\documentclass[useAMS]{mn2e}
\usepackage{latexsym,graphicx}

\usepackage{color}
\usepackage{amsmath}
\usepackage{bm}

\newcommand\kms{{\rm\,km\,s^{-1}}}
\newcommand\msun{\rm\,M_\odot}

\newcommand\lsun{\rm\,L_\odot}
\newcommand\hii{H\,{\sc ii} \,}

\newcommand\myr{\msun \, {\rm yr}^{-1}}

\title[An astrosphere around $\kappa$\,Cas]{An astrosphere around the blue supergiant $\kappa$\,Cas: possible explanation of its filamentary structure}

\author[O. A.~Katushkina et al.]{O. A.~Katushkina$^{1}$\thanks{E-mail: okat@iki.rssi.ru},
 D. B.~Alexashov$^{1,2}$, V. V.~Gvaramadze$^{1,3,4}$
 \newauthor and V. V.~Izmodenov$^{1,2,5}$ \\
$^{1}$Space Research Institute of Russian Academy of Sciences, Profsoyuznaya Str. 84/32, Moscow, 117335, Russia\\
$^{2}$Institute for Problems in Mechanics, prosp. Vernadskogo 101, block 1, Moscow, 119526, Russia\\
$^{3}$Sternberg Astronomical Institute, Lomonosov Moscow State University, Universitetskij Pr. 13, Moscow 119992, Russia\\
$^{4}$Isaac Newton Institute of Chile, Moscow Branch, Universitetskij Pr. 13, Moscow 119992, Russia \\
$^{5}$Lomonosov Moscow State University, GSP-1, Leninskie Gory, Moscow, 119991, Russia\\
}
\begin{document}

\def\vecv{\mbox{$\textbf{v}_p$}}
\def\vecV{\mbox{$\textbf{V}$}}
\def\vecB{\mbox{$\textbf{B}$}}
\def\vecq{\mbox{$\textbf{q}$}}
\def\vecQ{\mbox{$\textbf{Q}$}}

\date{Accepted 2017 September 15. Received 2017 September 6; in original form 2017 June 26.}

\pagerange{\pageref{firstpage}--\pageref{lastpage}} \pubyear{2017}

\maketitle

\label{firstpage}

\begin{abstract}
High-resolution mid-infrared observations carried out by the {\it Spitzer Space Telescope} allowed one
to resolve the fine structure of many astrospheres. In particular, they showed that the astrosphere around the
B0.7\,Ia star $\kappa$\,Cas (HD 2905) has a clear-cut arc structure with numerous cirrus-like
filaments beyond it. Previously, we suggested a physical mechanism for the formation of such filamentary structures.
Namely, we showed theoretically that they might represent the non-monotonic spatial distribution of the interstellar dust in
astrospheres (viewed as filaments) caused by interaction of the dust grains with the interstellar magnetic field
disturbed in the astrosphere due to colliding of the stellar and interstellar winds.
In this paper, we invoke this mechanism to explain the structure of the astrosphere around $\kappa$\,Cas.
We performed 3D magnetohydrodynamic modelling of the astrosphere for realistic parameters of the stellar
wind and space velocity. The dust dynamics and the density distribution in the astrosphere
were calculated in the framework of a kinetic model. It is found that the model results with the classical MRN
size distribution of dust in the interstellar medium do not match the observations, and that the observed
filamentary structure of the astrosphere can be reproduced only if the dust is composed mainly of big ($\mu$m-sized)
grains. Comparison of the model results with observations allowed us to estimate parameters
(number density and magnetic field strength) of the surrounding interstellar medium.
\end{abstract}

\begin{keywords} shock waves -- methods: numerical --
stars: individual: $\kappa$\,Cas -- (ISM:)dust, extinction.
\end{keywords}

\section{Introduction}
\label{sec:intro}

Interaction of the stellar wind with the circum- and interstellar medium (ISM) results in the formation of
structures called astrospheres. Severe interstellar extinction at low Galactic latitudes, where the majority of (massive) wind-blowing stars
is concentrated, makes the infrared (IR) observations the most effective way for detection and study of astrospheres (e.g. van Buren,
Noriega-Crespo \& Dgani 1995). Nowadays, with the advent of the {\it Spitzer Space Telescope}, {\it Wide-field Infrared Survey
Explorer} ({\it WISE}) and {\it Herschel Space Observatory}, many hundreds of new astrospheres were revealed (Peri et al. 2012;
Cox et al. 2012; Kobulnicky et al. 2016). Some of them have a distinct filamentary (cirrus-like)
structure (e.g. Gvaramadze et al. 2011a, b). Although the origin of this structure is not well understood, it seems likely that
the regular interstellar magnetic field might play an important role in its formation (Gvaramadze et al. 2011b).

Recently, Katushkina et al. (2017; hereafter Paper\,I) presented a physical mechanism possibly responsible for the origin of
the cirrus-like structure of astrospheres around runaway stars. It was shown that, under proper conditions, alternating minima
and maxima of the dust density (seen like filaments) might appear between the astrospheric bow shock (BS) and the astropause (AP)
because of periodical gyromotion of the dust grains around the interstellar magnetic field lines.

In the present work, we apply this mechanism to explain the morphology of the astrosphere around the
runaway blue supergiant $\kappa$\,Cas (HD\,2905). We choose this particular astrosphere because of its distinct cirrus-like
structure, as well as because the basic parameters of its associated star are known fairly well (see Section\,\ref{sec:obs}).
In Section\,\ref{sec:MHD}, we perform numerical modelling of interaction between the stellar wind and the magnetized ISM in
the framework of a 3D magnetohydrodynamic (MHD) model. In Section\,\ref{sec:kin}, we present the kinetic modelling of the
interstellar dust distribution in the wind-ISM interaction region. In Section\,\ref{sec:map}, we post-process the simulations
to make synthetic maps of infrared dust emission. In Section\,\ref{sec:com}, we compare the model results with observations,
estimate the interstellar plasma number density and magnetic field strength, and derive constraints on the dust parameters,
required to better reproduce the observations. Summary and discussion are presented in Section\,\ref{sec:dis}.

\section{Observational data}
\label{sec:obs}

\begin{figure}
\includegraphics[scale=0.4]{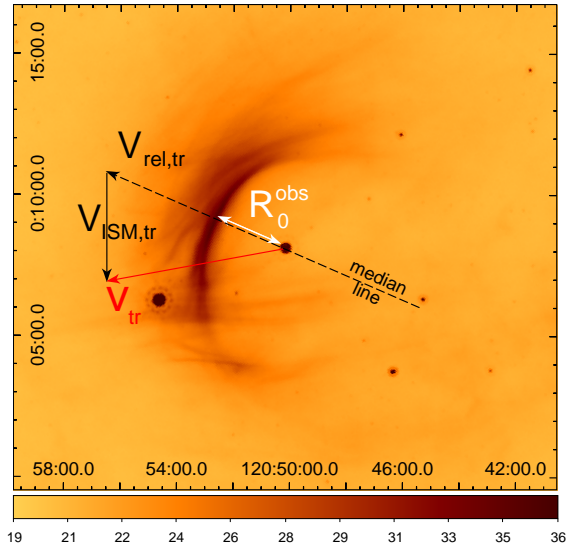}
\centering
\caption{{\it Spitzer} MIPS 24~$\mu$m image of the astrosphere around the
runaway blue supergiant $\kappa$\,Cas. The dashed line shows the median line (or symmetry line) of the astrosphere.
The red arrow shows the direction of transverse velocity of the star, $v_{\rm tr}$, as suggested
by the {\it Hipparcos} proper motion measurement. Transverse components of the ISM velocity vector
($v_{\rm ISM,tr}$) and the vector of the star's motion relative to the local ISM ($v_{\rm rel,tr}$)
are shown as well (see text for details). The minimum (projected) separation between $\kappa$\,Cas
and the brightest arc in the astrosphere is marked as $R_0^{\rm obs}$. The colour bar shows
the surface brightness on the linear scale in units of MJy sr$^{-1}$. The coordinates are the
Galactic longitude and latitude on the horizontal and vertical scales, respectively. At a distance
of 1 kpc, 1 arcmin corresponds to $\approx0.29$ pc.}
\label{fig:kapCas}
\end{figure}

The astrosphere around the blue supergiant (B0.7\,Ia; Walborn 1972) $\kappa$\,Cas
was discovered by van Buren \& McCray (1988) using the {\it Infrared Astronomical
Satellite} ({\it IRAS}) all-sky survey and presented for the first time in
van Buren, Noriega-Crespo \& Dgani (1995; see their fig.\,2c). In
the {\it IRAS} 60\,$\mu$m image, the astrosphere has an arc-like shape,
typical of bow shocks. The low resolution of
the {\it IRAS} data, however, did not allow to see fine details of
the astrosphere, which were revealed only with the advent of the
{\it Spitzer Space Telescope} and the {\it WISE} mission with their
much better angular resolution.

$\kappa$\,Cas was observed by {\it Spitzer} on 2007 September 18
(Program Id.: 30088, PI: A.Noriega-Crespo) using the Multiband Imaging Photometer
for {\it Spitzer} (MIPS; Rieke et al. 2004). We retrieved the post-basic data
calibrated MIPS 24\,$\mu$m image of $\kappa$\,Cas (with units of MJy\,${\rm
sr}^{-1}$) from the NASA/IPAC infrared science archive\footnote{http://irsa.ipac.caltech.edu/}.
In this image (presented for the first time
in Gvaramadze et al. 2011b) the astrosphere of $\kappa$\,Cas appears (see
Fig.~\ref{fig:kapCas}) as a clear arcuate structure with numerous cirrus-like
filaments beyond it, some of which are apparently attached to the main (brightest)
arc. The surface brightness of this arc is $\approx$35\,MJy\,${\rm sr}^{-1}$, while
that of the background is $\approx$20\,MJy\,${\rm sr}^{-1}$. Fig.~\ref{fig:kapCas}
also shows several filaments intersecting the arc at almost right angle to its surface
in the south part of the astrosphere. A possible origin of these filaments
is discussed in Section~\ref{sec:dis}.

We define the linear characteristic scale of the astrosphere as the minimum projected
distance between $\kappa$\,Cas and the brightest arc, $R_0^{\rm obs}$
(see Fig.~\ref{fig:kapCas}), which is related to the observed angular
separation between the star and the apex of the arc, $\Omega$, through the
relationship $R_0^{\rm obs}=\Omega d$, where $d$ is the heliocentric
distance to $\kappa$\,Cas. For $\Omega=2.6$ arcmin and $d=1$ kpc (see below),
one has $R_0^{\rm obs}\approx0.75$ pc or $2.3\times10^{18}$ cm.

It is believed that $\kappa$\,Cas belongs to the Cas\,OB14 association, which
is located at a distance of $d\approx1.0\pm0.1$ kpc
(Humphreys 1978; Mel'nik \& Dambis 2009). This distance is generally accepted in studies
of $\kappa$\,Cas (e.g. Crowther, Lennon \& Walborn 2006; Searle et al. 2008).
The presence of the astrosphere around $\kappa$\,Cas, however, suggests that this
star is a runaway and that it might be formed far away from its present position on the
sky (cf. Gvaramadze, Pflamm-Altenburg \& Kroupa 2011).
Correspondingly, $\kappa$\,Cas is not necessary a member of Cas\,OB14,
unless this star has obtained its peculiar space velocity because of dissolution of a
binary system in a recent supernova explosion in the association. In this connection,
we note that among four members of the association listed in Humphreys (1978)
one more star, HD\,2619 (B0.5\,III), produces a bow shock as well. The orientation
of this bow shock (visible in {\it WISE} 22 and 12\,$\mu$m images) suggests that HD\,2619
was injected in Cas\,OB14 from the open star cluster Berkeley\,59, located at a
distance of $\approx1$\,kpc (Pandey et al. 2008) and at $\approx3\fdg5$ (or $\approx60$\,
pc in projection) to the northwest from the star. It is possible therefore that Cas\,OB14
is actually a spurious association.

\begin{table}
  \caption{Basic parameters of $\kappa$\,Cas (Crowther et al. 2006; Searle et al. 2008).}
  \label{tab:sum}
  \begin{tabular}{ccccc}
\hline
   $v_\infty$ & $\dot{M}$ & $T_{*}$ & $R_*$ & $\log(L_*/\lsun)$ \\
  ($\kms$) & ($\myr$) & (kK) & ($R_\odot$) &  \\
\hline
  850$-$1000 & $(2.0-2.5)\times10^{-6}$ & 23.5 & 33 & 5.48 \\
\hline
\end{tabular}
\end{table}

\begin{table*}
\caption{Summary of astrometric and kinematic data on
$\kappa$\,Cas (see text for details).}
\label{tab:prop}
\begin{tabular}{ccccccccc}
\hline
$d$ & $\mu _\alpha \cos \delta$
& $\mu _\delta$ & $v_{\rm r,hel}$ & $v_{\rm l}$ & $v_{\rm b}$ & $v_{\rm r}$ & $V_{\rm tr}$ & $V_*$ \\
(kpc) & (mas ${\rm yr}^{-1}$) & (mas ${\rm yr}^{-1}$) & ($\kms$) & ($\kms$) & ($\kms$) & ($\kms$) & ($\kms$) & ($\kms$) \\
\hline
1.0 & 3.65$\pm$0.17 & $-$2.07$\pm$0.16 & 0.3$\pm$0.8 & 20.8$\pm$0.8 & $-$3.8$\pm$0.8 & 18.5$\pm$0.8 & 21.1$\pm$1.1 & 28.1$\pm$1.4 \\
\hline
\end{tabular}
\end{table*}

A somewhat larger distance to $\kappa$\,Cas follows from the {\it Hipparcos} parallax
(van Leeuwen 2007) and the empirical relationship between the strength of the interstellar
Ca\,{\sc ii} lines and the distances to early-type stars (Megier et al. 2009), yielding
respectively $d=1.37^{+0.42} _{-0.25}$ and $d=1.46\pm0.30$ kpc. Taken at face value, these
two distance estimates imply a too high bolometric luminosity of $\log(L_*/\lsun)\approx5.8$,
which along with the effective temperature of $\kappa$\,Cas of $T_*=23.5\pm1.5$\,kK (Searle et
al. 2008) would place this star on the S\,Doradus instability strip (Wolf 1989) in the
Hertzsprung-Russell diagram. Since $\kappa$\,Cas does not show variability typical of stars in
this region of the Hertzsprung-Russell diagram, it is likely that it is located at a shorter
distance. In what follows, we adopt the distance to $\kappa$\,Cas of d=1~kpc. The basic parameters
of $\kappa$\,Cas ($T_*$, $L_*$, stellar wind velocity $v_\infty$, mass loss rate $\dot{M}$
and radius $R_*$) are compiled in Table\,\ref{tab:sum}.

In Table\,\ref{tab:prop} we provide astrometric and kinematic data
on $\kappa$\,Cas. The proper motion measurements, $\mu _\alpha \cos
\delta$ and $\mu _\delta$, are based on the new reduction of the
{\it Hipparcos} data by van Leeuwen (2007). The heliocentric
radial velocity of the star, $v_{\rm r,hel}$, is taken from Gontcharov (2006).
Using these data, the Solar galactocentric distance $R_0 = 8.0$ kpc and the
circular Galactic rotation velocity $\Theta _0 =240 \, \kms$ (Reid et al. 2009),
and the solar peculiar motion $(U_{\odot},V_{\odot},W_{\odot})=(11.1,12.2,7.3) \,
\kms$ (Sch\"onrich, Binney \& Dehnen 2010), we calculated the peculiar transverse
velocity $V_{\rm tr}=(v_{\rm l}^2 +v_{\rm b}^2)^{1/2}$, where $v_{\rm l}$ and
$v_{\rm b}$ are, respectively, the velocity components along the Galactic longitude
and latitude, the peculiar radial velocity $v_{\rm r}$, and the total space
velocity $V_*$ of the star. For the error calculation,
only the errors of the proper motion and the radial velocity measurements were
considered. The obtained space velocity of $\approx30 \, \kms$ implies that
$\kappa$\,Cas is a classical runaway star (e.g. Blaauw 1961).

The orientation of the symmetry axis of astrospheres around moving stars is determined by
the orientation of the stellar velocity relative to the local ISM. In a static,
homogeneous ISM and for a spherically-symmetric stellar wind, the symmetry axis of an
astrosphere is aligned with the vector of stellar space motion. In this case, the geometry
of detected astrospheres (bow shocks) can be used to infer the direction of stellar motion
and thereby to determine possible parent clusters for the bow-shock-producing stars (e.g.
Gvaramadze \& Bomans 2008). In reality, however, the ISM might not necessary be at rest owing
to the effects of nearby supernova explosions, expanding \hii regions, or outflows from massive
star clusters. Also, the shape of astrospheres of hot (runaway) stars might be affected
by photoevaporation flows from nearby regions of enhanced density (cloudlets) caused by
ultraviolet emission of these stars (e.g. Mackey et al. 2015; Gvaramadze et al. 2017).

Fig.\,\ref{fig:kapCas} shows that the vector of the peculiar (transverse) velocity of
$\kappa$\,Cas is misaligned with the symmetry axis (median line) of the astrosphere by an angle
$\alpha\approx35\degr$. This misalignment might be caused by inaccuracy of the space
velocity calculation\footnote{Note that the
misalignment would be even stronger if $\kappa$\,Cas is located at $d>1$ kpc.} or by the
presence of a regular flow in the local ISM. We consider the latter possibility to be more
likely because the {\it Hipparcos} proper motion measurement for $\kappa$\,Cas is very
reliable (see Table\,\ref{tab:prop}). We suggest, therefore, that the orientation of the
astrosphere around $\kappa$\,Cas is affected by a flow of the local ISM.

To reconcile the orientation of the astrosphere around $\kappa$\,Cas with the orientation of the
stellar transverse motion, one needs to assume that the ISM is moving in the north-south
direction (i.e. almost along the Galactic longitude) with a transverse velocity of $V_{\rm ISM,tr}\approx15 \, \kms$.
In this case, the transverse component of the relative velocity between the star and the ISM is
$V_{\rm rel,tr}\approx26 \, \kms$. Allowing the possibility that the ISM could have a velocity component
in the radial direction as large as in the transverse one, i.e. $\pm15 \, \kms$, one finds that the total
relative velocity, $V_{\rm rel}$, could range from 26 to $42 \, \kms$. For the sake of certainty, in our
calculations we adopt an intermediate value of $V_{\rm rel}$ of $35 \, \kms$, which corresponds
to the angle between the vector of the total relative velocity and the line of sight of $\theta=48\degr$.
Note that the orientation of the astrosphere would not change if the transverse velocity of the ISM has
a component in the east-west direction as well (i.e. parallel to the Galactic plane). The actual value
of $V_{\rm rel}$ (or the sonic Mach number), however, is not critically important for our calculations
because in the considered case (see below) the overall structure of the astrosphere is mostly determined
by the interstellar magnetic field.

\section{Numerical model}
\label{sec:num}

To compare the model astrospheres with observations one needs to produce synthetic maps of the thermal emission from
the dust accumulated at the region of interaction between the stellar wind and the surrounding ISM.
For this, the following steps should be performed: 1) MHD modelling of the plasma and magnetic
field distributions in the astrosphere, 2) kinetic modelling of the interstellar dust distribution in the wind-ISM
interaction region, and 3) determination of the dust temperature and calculation of the thermal emission intensity. These
steps are presented in the next subsections.

\subsection{3D MHD modelling of the astrosphere}
\label{sec:MHD}

\begin{figure}
\includegraphics[scale=0.5]{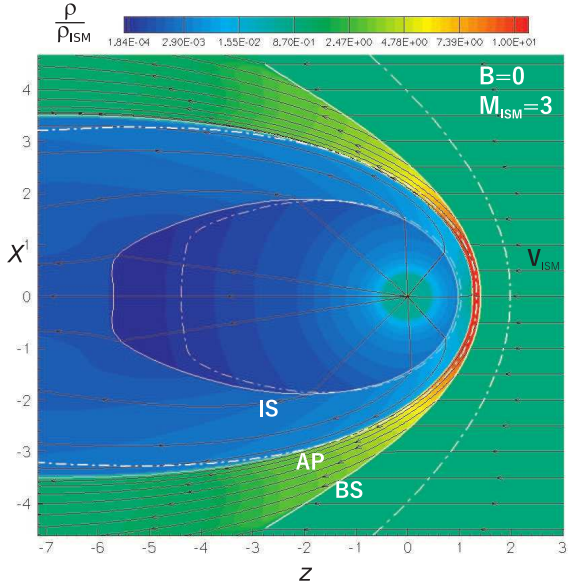}
\centering
\caption{An axisymmetric model astrosphere in the ISM without magnetic field.
The sonic Mach number is equal to 3. Solid and dot-dashed white curves show the positions of discontinuities
(IS -- inner shock, AP -- astropause, BS -- bow shock) for models with and without radiative cooling, respectively.
The distribution of the plasma density in the model with cooling is colour-coded. Streamlines are shown by black lines.
}
\label{fig:2D_cool}
\end{figure}

\begin{figure}
\includegraphics[scale=0.5]{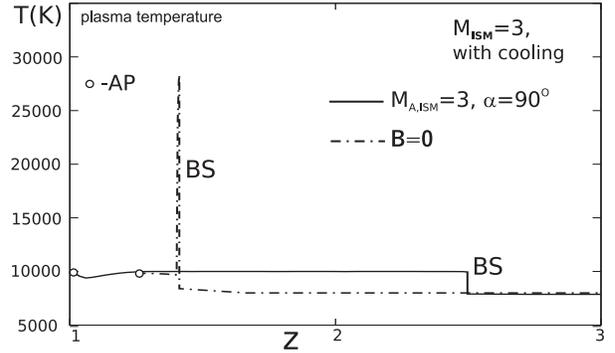}
\centering
\caption{Plasma temperature distribution along the $Z$-axis (directed in the opposite direction to the ISM flow)
outside of the astropause (AP) for model astrospheres with (solid lines) and without (dot-dashed lines)
perpendicular magnetic field. In both models the radiative cooling is included and $M_{\rm ISM}$=3. BS stands
for bow shock.}
\label{fig:1D_cool}
\end{figure}

To determine the distribution of plasma and magnetic field in and around the astrosphere we use the steady-state 3D MHD
model described in Paper\,I. Here we also take into account the radiative cooling of plasma with the cooling function from
Cowie et al. (1981). Note that the interstellar magnetic field is essential for this study due to two reasons. Firstly,
the proposed physical mechanism for formation of filaments is based on gyrorotation of dust grains around magnetic field
lines frozen into the interstellar plasma. Secondarily, in the absence of the magnetic field the radiative cooling
strongly reduces the thickness of the layer between the AP and the BS because of loss of energy (see Fig.~\ref{fig:2D_cool}).
Therefore, the filaments formed in this region would be unresolvable. In the presence of significant interstellar magnetic
field the outer shock layer does not collapse (see Fig.~\ref{fig:1D_cool}) and the filaments might be observable.

Note that hereafter all distances in plots are dimensionless, they are normalized to the stand-off distance, which is the
distance from the star to the AP in the upwind direction in the unmagnetized medium:
\begin{equation}
D^*=\sqrt{\frac{\dot{M} v_\infty}{4\pi \rho_{\rm p,ISM}
V_{\rm rel} ^2}},
\label{eqn:D}
\end{equation}
where $\rho_{\rm p,ISM}= 1.4n_{\rm p,ISM}\,m_{\rm p}$, $n_{\rm p,ISM}$ is the ISM number density of protons
and $m_{\rm p}$ is the proton mass.

In the dimensionless form the solution of the problem depends on the sonic ($M_{\rm ISM}$)
and Alfvenic ($M_{\rm A,ISM}$) Mach numbers of the ISM. In general, there is one more dimensionless parameter
characterizing the efficiency of the radiative cooling. However, in the case of relatively high
ISM number density ($n_{\rm p,ISM}\geq2$~cm$^{-3}$, see~Section~\ref{sec:com}) and for the adopted cooling
function the plasma temperature in the outer shock region (i.e. between the AP and the BS) remains almost
constant (see the solid curve in Fig.~\ref{fig:1D_cool}) and, correspondingly, the results do not depend on
the dimensionless parameter related to the radiative cooling.

We perform calculations with the ``perpendicular" interstellar magnetic field, i.e. with the magnetic field
vector in the undisturbed ISM (${\boldsymbol B}_{\rm ISM}$) perpendicular to the relative ISM velocity vector
(${\boldsymbol V}_{\rm ISM}=-{\boldsymbol V}_{\rm rel}$). We choose this orientation of the
magnetic field because the nose part of the astrosphere around $\kappa$\,Cas seems to be axisymmetric
and because we know from Paper\,I that in the case of parallel magnetic field the dust is accumulated in
filaments at flanks of the astrosphere and is absent in its nose part (see fig.\,10 in Paper\,I). The effect
of the stellar magnetic field is neglected in our calculations.

We assume that the ISM temperature is $\approx7000-8000$\,K, which along with $V_{\rm rel}=35 \, \kms$ (see Section\,\ref{sec:obs})
corresponds to the sonic Mach number of $M_{\rm ISM}\approx3$. We performed calculations
for several values of $M_{\rm A,ISM}$. The magnitude of the interstellar magnetic field (and Alfvenic Mach number) determines the thickness of the outer shock layer. It is found that the best qualitative agreement between the model results and the observations could be achieved for $M_{\rm A,ISM}\approx1.5-3$ (see discussion in Section\,\ref{sec:dis}). The main part of the calculations presented in this work is performed for $M_{\rm A,ISM}=1.77$.

\begin{figure*}
\includegraphics[scale=0.8]{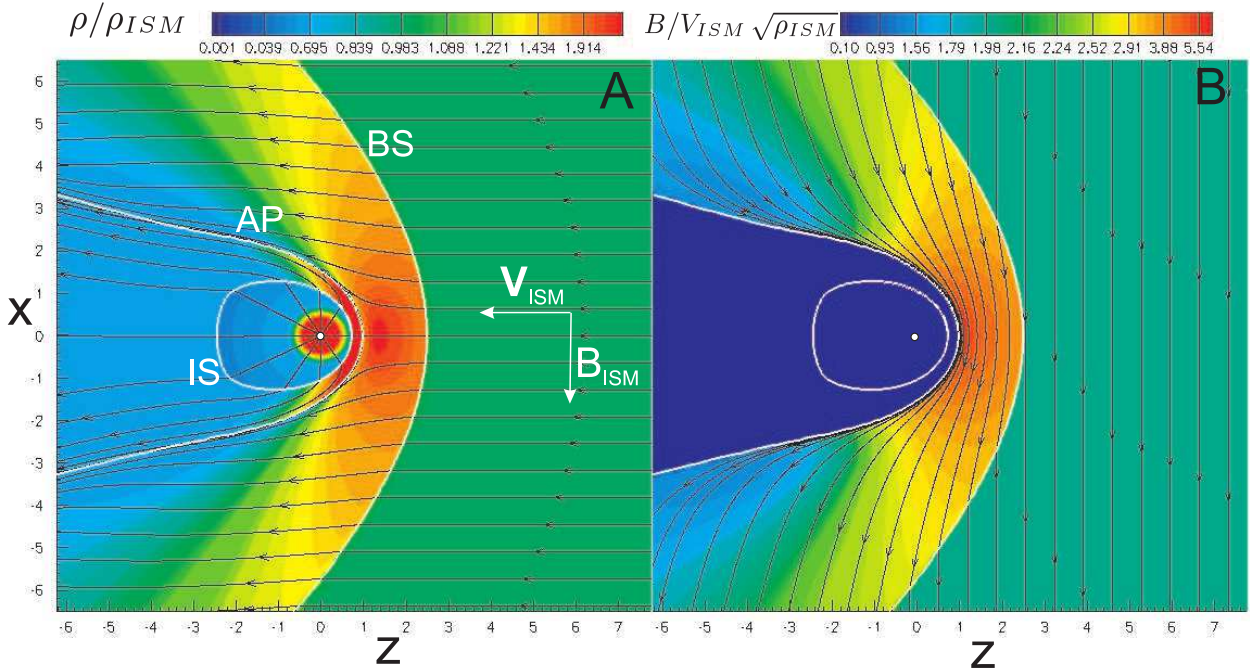}
\centering
\caption{2D distributions of the plasma density (panel A) and the magnetic field (panel B) in the $({\boldsymbol
V}_{\rm ISM},{\boldsymbol B}_{\rm ISM})$-plane. The inner shock (IS), the astropause (AP) and the bow shock (BS)
are plotted with white lines. The streamlines (panel A) and the magnetic field lines (panel B) are plotted with
black lines.}
\label{fig:plasma}
\end{figure*}

Fig.\,\ref{fig:plasma} plots 2D distributions of the plasma density and the magnetic field in the
$({\boldsymbol V}_{\rm ISM},{\boldsymbol B}_{\rm ISM})$-plane.
Note that the thickness of the outer shock layer between the AP and the BS is determined by the chosen Alfvenic Mach number,
i.e. the weaker the magnetic field (or the larger $M_{\rm A,ISM}$) the thinner the layer.

\subsection{Kinetic modelling of the dust distribution in the astrosphere}
\label{sec:kin}

To calculate the dust distribution in the astrosphere we use the kinetic model described in Alexashov et al. (2016) and
Paper\,I. In general, the dynamics of charged interstellar dust grains in the astrosphere is determined by the following forces:
the electromagnetic Lorentz force ${\boldsymbol F}_{\rm L}$, the stellar gravitational force ${\boldsymbol F}_{\rm G}$, the
stellar radiation pressure ${\boldsymbol F}_{\rm rad}$, and the drag force ${\boldsymbol F}_{\rm drag}$ due to interaction of the dust grains
with protons and electrons through the direct and Coulomb collisions. Therefore, the motion equation of a charged dust grain
is the following:
\begin{eqnarray}
{{\rm d}{\boldsymbol v}_{\rm d} \over {\rm d}t}={q\over m_d c}({\boldsymbol v}_{\rm rel}\times {\boldsymbol B})+\left(-GM_*
+ {\sigma_{\rm d}\bar{Q}_{\rm rp}L_{*} \over 4\pi m_d c}\right){{\boldsymbol e}_{\rm r} \over r^2} \nonumber \\
 + \sigma_{\rm d} n_{\rm p} kT_p \hat{G}(v_{\rm rel}, T_p){{\boldsymbol v}_{\rm rel} \over |{\boldsymbol v}_{\rm rel}|},
\label{eq:motion}
\end{eqnarray}
where $q$ and $m_d$ are the charge and mass of the grain, $c$ is the speed of light,
${\boldsymbol v}_d=\dot{{\boldsymbol x}}_{\rm d}$, ${\boldsymbol x}_{\rm d}$ is the position of a dust grain,
$r=|{\boldsymbol x}_{\rm d}|$
is the distance from the star, ${\boldsymbol e}_{\rm r}$ is the unit vector in the radial direction,
${\boldsymbol v}_{\rm p}$ is the local plasma velocity,
${\boldsymbol v}_{\rm rel}={\boldsymbol v}_{\rm d}-{\boldsymbol v}_{\rm p}$
is the relative velocity between the plasma and the dust grain, ${\boldsymbol B}$ is the magnetic field, $G$ is the gravitational
constant, $M_*$ and $L_*$ are the stellar mass and luminosity, $\sigma_{\rm d}=\pi r_{\rm d} ^2$ is the geometrical
cross section of the dust grain, $r_{\rm d}$ is the radius of the grain, $\bar{Q}_{\rm rp}\approx1$ is the flux-weighted
mean radiation pressure efficiency of the grain, $n_{\rm p}$ and $T_p$ are the plasma
number density and temperature, $k$ is the Boltzmann constant, and $\hat{G}(v_{\rm rel}, T)$ is the dimensionless function
determining the drag force (see, e.g., Draine \& Salpeter 1979; Ochsendorf et al. 2014).
The plasma parameters ($n_{\rm p}$, ${\boldsymbol v}_{\rm p}$ and ${\boldsymbol B}$) are taken from the MHD model
described above.

It is assumed that the dust grain charge $q$ is constant along the trajectory and $q=U_{\rm d,ISM}\,r_{\rm d}$,
where $U_{\rm d,ISM}=+0.75$~V is the dust surface potential commonly assumed for the local ISM around the Sun (Gr\"{u}n \& Svestka 1996). The potential is
positive due to an influence of accretion of protons, photoelectric emission of stellar and interstellar radiation, and
secondary electron emission (see, e.g. Kimura \& Mann 1998; Akimkin et al. 2015).

We estimated the relative contribution of the listed forces to the dust dynamics in an astrosphere of
$\kappa$\,Cas and found that the stellar gravitation attractive force and the drag force are negligible compared with others.
Stellar radiation force is especially important for small dust grains with radii $r_{\rm d}\leq0.1 \,\mu$m. These dust
grains are swept out far away by the stellar radiation and do not cross the BS.
Therefore, in our calculations we consider the dust grains with $r_{\rm d}>0.2 \, \mu$m.

Classical power-law MRN size distribution (Mathis, Rumpl \& Nordsieck, 1977) is assumed for the dust grains in the undisturbed ISM, $n_{\rm d,ISM}(r_{\rm d})\sim r_{\rm d}^{-3.5}$, with the minimum and maximum
dust grain radii of 0.2 and $3\,\mu$m, respectively. Corresponding mass of the dust grains ranges from $8.37\times10^{-14}$
to $2.83\times10^{-10}$\,g (it is assumed that the dust grains are spherical and have a uniform density $\rho_{\rm d}=2.5 \, {\rm g} \,
{\rm cm}^{-3}$).
It is also assumed that in the undisturbed ISM all dust grains have the same velocity of ${\boldsymbol V}_{\rm ISM}$. The kinetic equation~(\ref{eq:motion}) is solved by the imitative Monte-Carlo method.

\begin{figure*}
\includegraphics[scale=0.8]{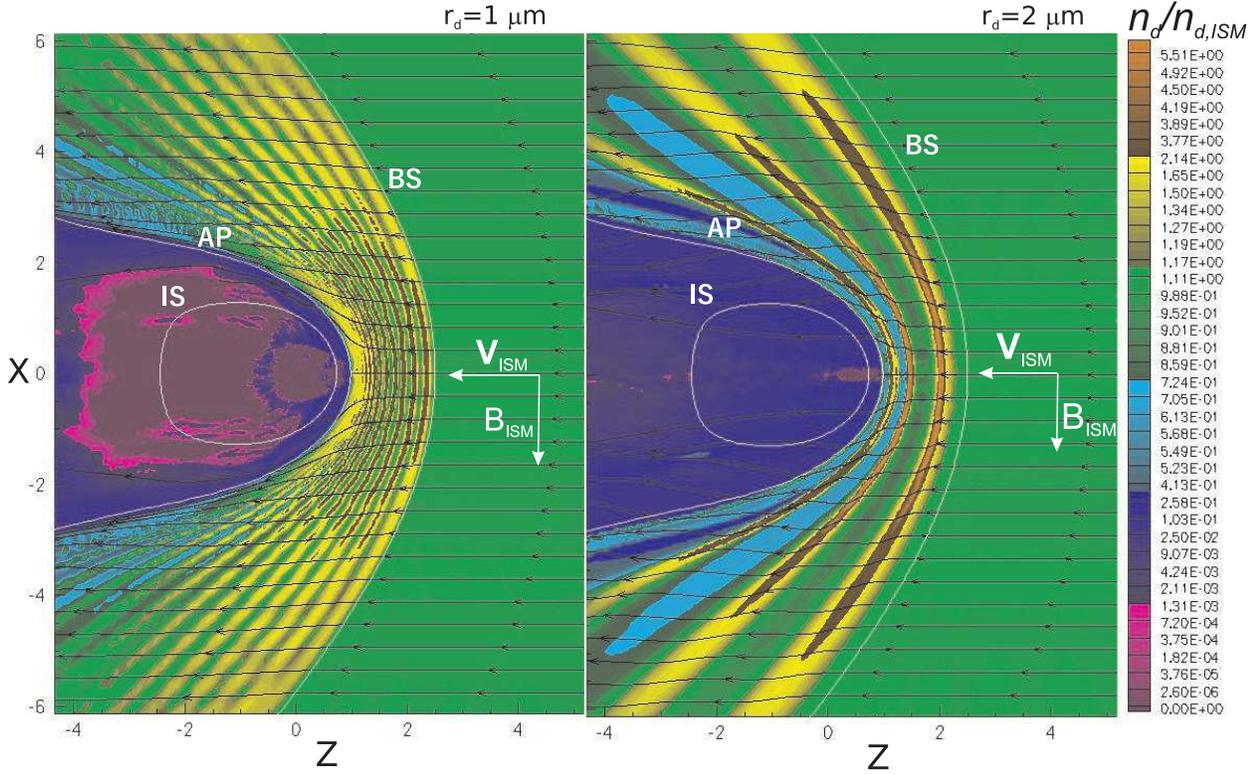}
\centering
\caption{2D distribution of the dust number density in the $({\boldsymbol B}_{\rm ISM},{\boldsymbol V}_{\rm ISM})$-plane for dust
grains with $r_{\rm d}=1$ and $2\,\mu$m. The inner shock (IS), the astropause (AP) and the bow shock (BS) are shown
by white lines. The streamlines (panel A) and the magnetic field lines (panel B) are plotted with
black lines.} \label{fig:ndust}
\end{figure*}

Fig.~\ref{fig:ndust} plots the distribution of the dust number density in the
(${\boldsymbol B}_{\rm ISM},{\boldsymbol V}_{\rm ISM})$-plane for dust grains with
$r_{\rm d}=1\,\mu$m and $2\,\mu$m. In both cases, the filaments are clearly seen. Their
formation is caused by periodical gyrorotation of charged dust
grains around the magnetic field lines. This physical mechanism is extensively discussed in Paper\,I. The filaments produced by larger dust
grains are sparser because of the larger gyroradius.
It is shown in Paper\,I that the characteristic separation between filaments is:
\begin{equation}
D_{\rm gyr}=v_{{\rm p},z}\frac{2\pi m_d}{B q}=v_{{\rm p},z}\frac{8 \pi^2 \rho_{\rm d}}{3}\frac{r_{\rm d} ^2}{B \, U_{\rm {d,ISM}}},
\label{eqn:Dgyr}
\end{equation}
where $v_{{\rm p},z}$ is the component of the plasma velocity along the $Z$-axis.
For small $D_{\rm gyr}$ the filaments merge with each other and cannot be resolved. For the chosen model
parameters ($M_{\rm A,ISM}$, $M_{\rm ISM}$, $\rho_{\rm d}$ and $U_{\rm ISM}$) the distinct filamentary structure is
formed for $r_{\rm d}\approx1.5-3 \, \mu$m. From equation\,(\ref{eqn:Dgyr}) it follows that for smaller values of
$B$ and/or $U_{\rm ISM}$ the filamentary structure of the astrosphere would be discernible if the size of the
dust grains would be reduced accordingly.

\subsection{Synthetic maps of thermal dust emission}
\label{sec:map}

The astrosphere around $\kappa$\,Cas, like the majority of other known astrospheres, is visible only via its infrared emission,
whose origin can be attributed mostly to the thermal dust emission (e.g. van Buren \& McCray 1988). To compare the model
astrosphere with the {\it Spitzer} MIPS image we produce a synthetic map of thermal dust emission at $24\,\mu$m.
For this, we integrate the local emissivity over the line of sight:
$I_{\nu}(r_{\rm d})=\int j_{\rm \nu}(r_{{\rm d}},s)ds$, where $s$ is the coordinate along this line.

\begin{figure*}
\includegraphics[scale=0.8]{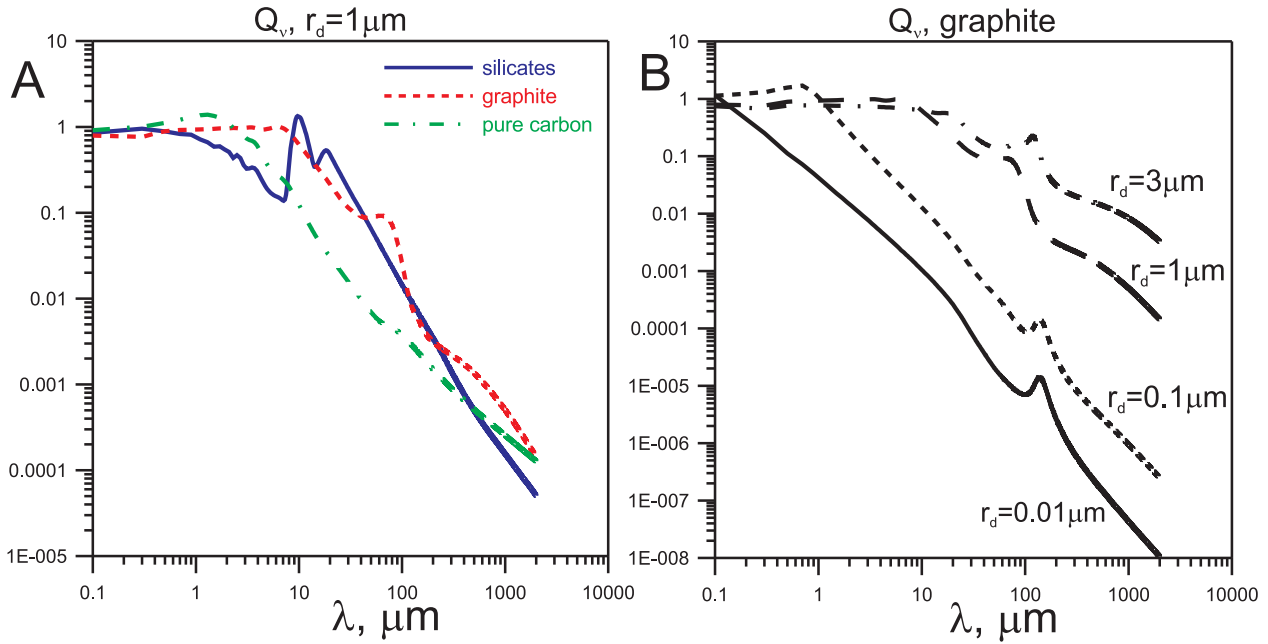}
\centering
\caption{Dimensionless efficiency absorption factor of dust as a function of wavelength.
A. $Q_{\nu}$ is presented for a fixed grain radius of 1~$\mu$m and different materials. B. $Q_{\nu}$ is presented
for graphite grains with different radii.} \label{fig:Qnu}
\end{figure*}

\begin{figure*}
\includegraphics[scale=0.8]{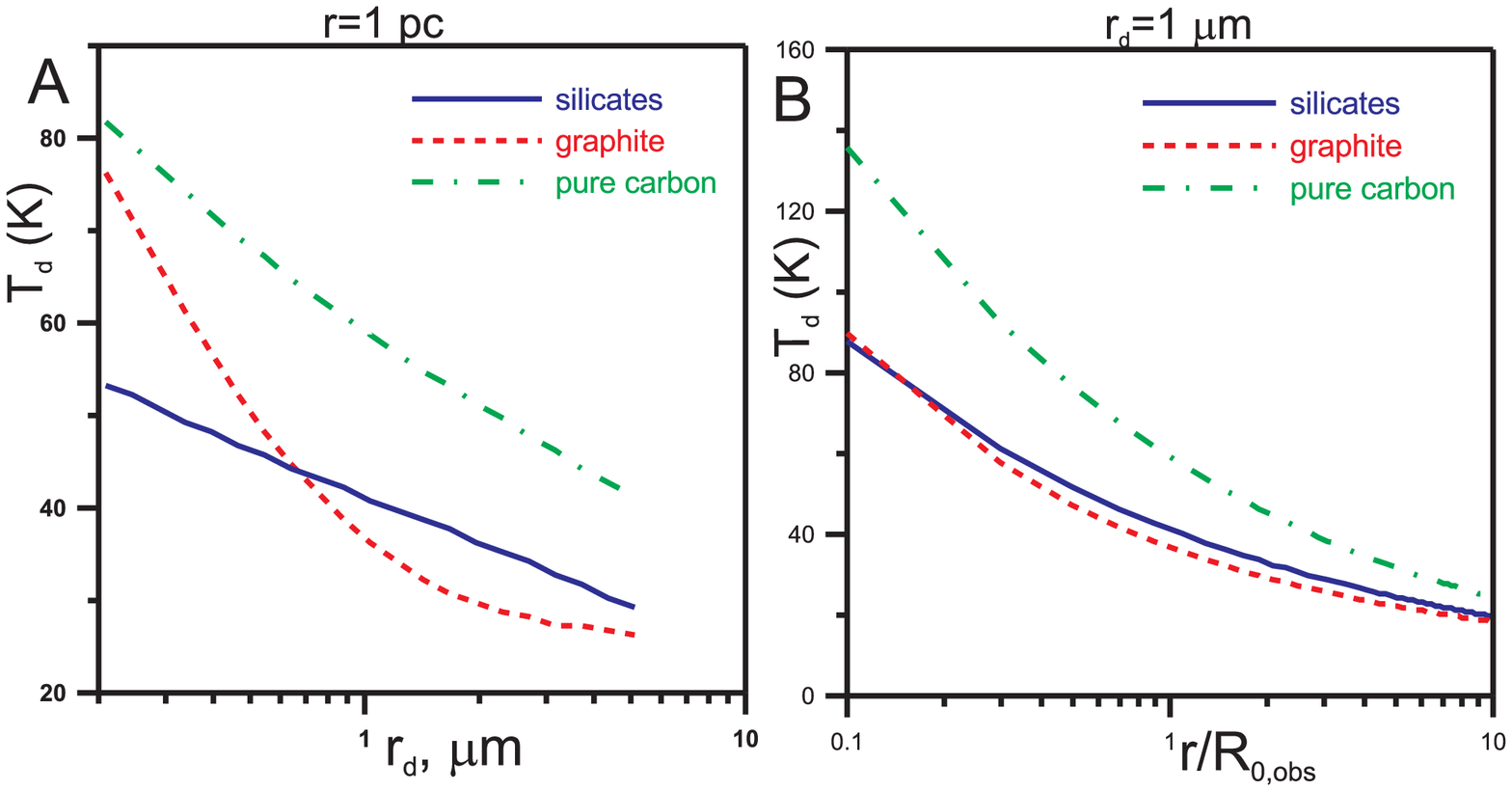}
\centering
\caption{A. Dust temperature at a distance $r=1$~pc from the star as a function of the dust grain radius $r_{\rm d}$.
B. Temperature of the dust grains with $r_{\rm d}=1\, \mu$m as a function of distance.}
\label{fig:Td}
\end{figure*}

The local emissivity can be expressed as follows:
\begin{equation*}
j_\nu(r_{\rm d},s)=\pi r_{\rm d}^2 \, n_{\rm d}(s,r_{\rm d}) Q_\nu (r_{\rm d}) B_\nu (T_{\rm d}(s,r_{\rm d})),
\end{equation*}
where $n_{\rm d}$ is the dust number density, $Q_\nu$ is the dimensionless efficiency absorption factor of dust,
$B_\nu$ is the Planck function, $T_{\rm d}$ is the dust temperature.
$Q_{\nu}(r_{\rm d})$ is calculated using the Mie theory (Bohren \&
Huffman, 1983) for the chosen dust material. In this work, we consider astronomical silicates (Draine 2003), graphite
(Draine 2003) and pure carbon (J\"ager, Mutschke \& Henning 1998). Plots of $Q_{\nu}(r_d)$ for different materials
and grain radii are presented in Fig.~\ref{fig:Qnu}.
The dust temperature can be calculated from local energy balance between dust heating by the stellar radiation and
cooling due to thermal emission.
Details of the temperature calculations are given in Appendix~A.
The obtained temperature depends on the grain material and radius, and the distance from the star (see Fig.~\ref{fig:Td}).

Dust number density can be represented as $n_{\rm d}(s,r_{\rm d})=\hat{n}_d(s,r_{\rm d})n_{\rm d,ISM}(r_{\rm d})$, where
$\hat{n}_{\rm d}$ is the dimensionless dust number density taken from the model results. In the case of the MRN size distribution,
the ISM number density of dust grains with radii in the range $[r_{\rm d}-{\rm d}r_{\rm d}/2; r_{\rm d}+{\rm d}r_{\rm d}/2]$ is
${\rm d}n_{\rm d,ISM}(r_{\rm d})=N_{\rm ISM}r_{\rm d}^{-3.5} d r_{\rm d}$. The dimensional coefficient $N_{\rm ISM}$ can be found from the assumption that
the typical gas to dust mass ratio is equal to 100 (see Appendix~B).
Thus, the local emissivity of the dust grains with radii in the above range is
\begin{equation*}
{\rm d}j_{\nu}(s,r_{\rm d})=\pi r_{\rm d}^2 \, Q_{\nu}(r_{\rm d}) \, B_{\nu}(T_{\rm d}(s,r_{\rm d})) \,
\hat{n}_{\rm d}(s,r_{\rm d})N_{\rm ISM}r_{\rm d}^{-3.5} \, {\rm d}r_{\rm d}
\end{equation*}
and the total intensity is:
\begin{eqnarray}
I_{\nu}=\pi \, N_{\rm ISM}\, \int_{r_{\rm d,min}}^{r_{\rm d,max}}{\rm d}r_{\rm d} \int  r_{\rm d}^{-1.5} \, Q_{\nu}(r_{\rm d})
\, B_{\nu}(T_{\rm d}(s,r_{\rm d})) \nonumber \\
\times \hat{n}_{\rm d}(s,r_{\rm d}) \, ds. \nonumber
\end{eqnarray}
We also performed calculations for certain dust grain radii with a uniform dust distribution in a narrow range
$[r_{{\rm d},0}- dr_{\rm d}/2, r_{{\rm d},0}+dr_{\rm d}/2]$.
In this case, in the formula for the total intensity $r_{\rm d}^{-1.5}$ should be replaced with $r_{{\rm d},0}^2$.

\section{Comparison of the model results with observations and evaluation of the ISM parameters}
\label{sec:com}

\begin{figure}
\includegraphics[width=8cm]{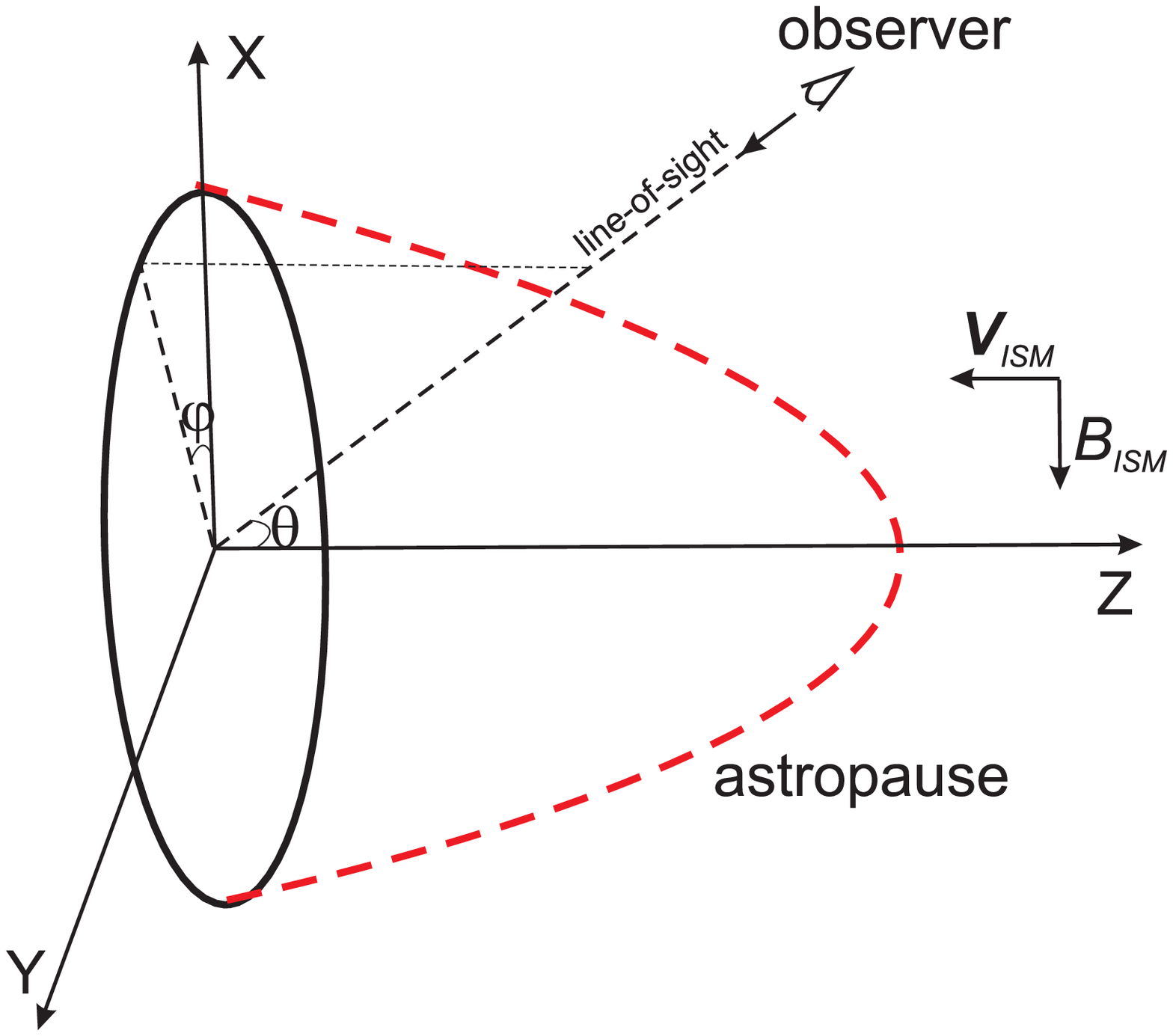}
\centering
\caption{Schematic representation of observation of an astrosphere
from an arbitrary line of sight. The astropause is shown by a red dashed
line. The line of sight is characterized by two spherical angles:
$\theta$ and $\phi$.} \label{fig:scheme}
\end{figure}

To compare the model results with the observational data the intensity of the thermal dust emission
should be calculated in the plane perpendicular to the line of sight. The orientation of
the stellar velocity with respect to the line of sight is determined by two spherical angles $\theta$ and $\phi$
(see Fig.~\ref{fig:scheme} for illustration). In our calculations the angle $\theta$ is fixed at 48$\degr$
(see Section\,\ref{sec:obs}), while the angle $\phi$ is a free parameter of the model because it is determined by the
unknown orientation of the $({\boldsymbol B}_{\rm ISM},{\boldsymbol V}_{\rm ISM})$-plane. Therefore, we performed
calculations at different planes and found that $\phi\approx110-150\degr$ provides the best agreement with the
observations. The results are presented for intermediate $\phi\approx135\degr$.

\begin{figure}
\includegraphics[width=8cm]{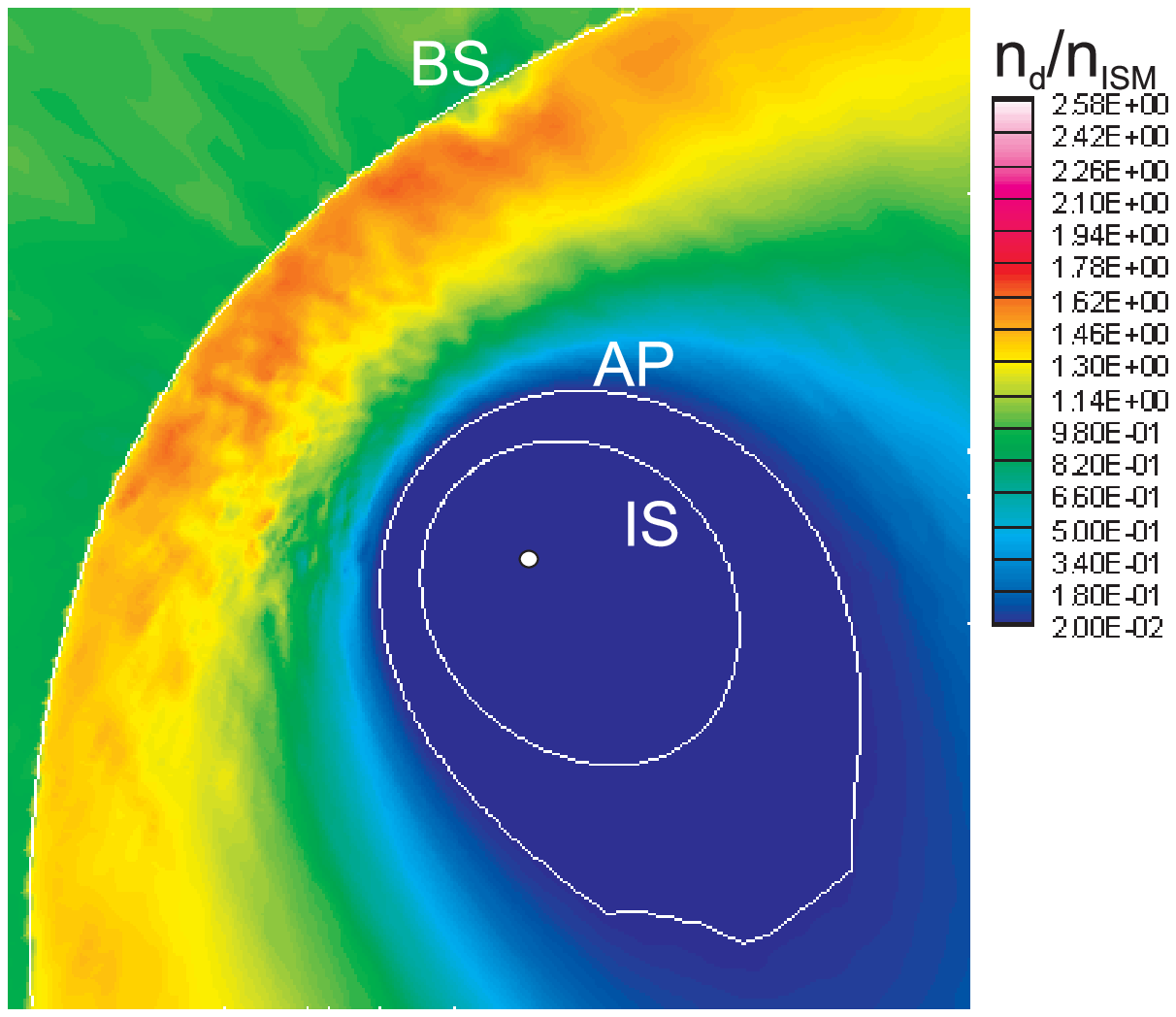}
\centering
\caption{Dust number density in the observational plane for grains with the MRN size distribution in the
range of $0.2-3\, \mu$m.}
\label{fig:n_tot}
\end{figure}

Fig.~\ref{fig:n_tot} plots the dust number density obtained for the MRN size distribution in the ISM with
the range of grain sizes of $r_{\rm d}=0.2-3\, \mu$m. It is seen that no filaments are visible. The reason
for this is that the filaments formed by dust grain with continuous size distribution merge with each other
in a wide arcuate structure between the BS and the AP.

Before considering the intensity maps, we note that all our model calculations were performed in dimensionless form.
In order to transform the dimensionless solution in the dimensional form and find absolute values of intensities one needs
to specify the required dimensional parameters of the model (e.g. the characteristic distances and the dust number density
in the ISM). This can be done in the following way. We assume that the brightest arc in the astrosphere around $\kappa$\,Cas
coincides with the astropause (our numerical calculations below support this assumption). By comparison of the obtained
dimensionless solution with the known distance to the brightest arc, $R_0^{\rm obs}$, it is possible to determine the
characteristic distance $D^*$, the corresponding ISM number density and the magnetic field strength, which are consistent
with the observations and the model results. Namely, from the numerical modelling we know the dimensionless distance from
the star to the astropause in the nose part of the astrosphere, $\hat{R_0}$, and:
\begin{equation*}
R_0^{\rm obs}=\hat{R_0}D^*.
\end{equation*}
Combining this equation with equation\,(\ref{eqn:D}) one has:
\begin{equation*}
n_{\rm p,ISM}=\frac{\dot{M} v_\infty}{5.6\pi m_{\rm p}V_{\rm rel}^2}\left(\frac{\hat{R_0}}{R_0^{\rm obs}}\right)^2.
\end{equation*}
Then we obtain $B_{\rm ISM}$ from the Alfvenic Mach number:
\begin{equation*}
B_{\rm ISM}=\frac{(5.6\pi m_{\rm p}n_{p, \rm ISM})^{1/2} V_{\rm rel}}{M_{\rm A,ISM}}.
\end{equation*}
The following estimates are obtained: $n_{\rm p,ISM}=3-11$\,cm$^{-3}$, $B_{\rm ISM}=18-35 \, \mu$\,G.
The ranges of $n_{\rm p,ISM}$ and $B_{\rm ISM}$ are due to uncertainties in $\dot{M}$ and $v_\infty$ of
$\kappa$\,Cas (see Table~\ref{tab:sum}). From the gas to dust mass ratio of 100, one obtains the dust number density
and the constant $N_{\rm ISM}$ (see Appendix~B). Note that the intensity of the thermal dust emission is proportional to
$N_{ISM}$, which in turn is proportional to $n_{\rm p,ISM}$. All intensity maps presented below are computed for
the intermediate value of $n_{\rm p,ISM}=5$~cm$^{-3}$ and the corresponding uncertainties in intensity are a factor
of few.

\begin{figure*}
\includegraphics[scale=0.8]{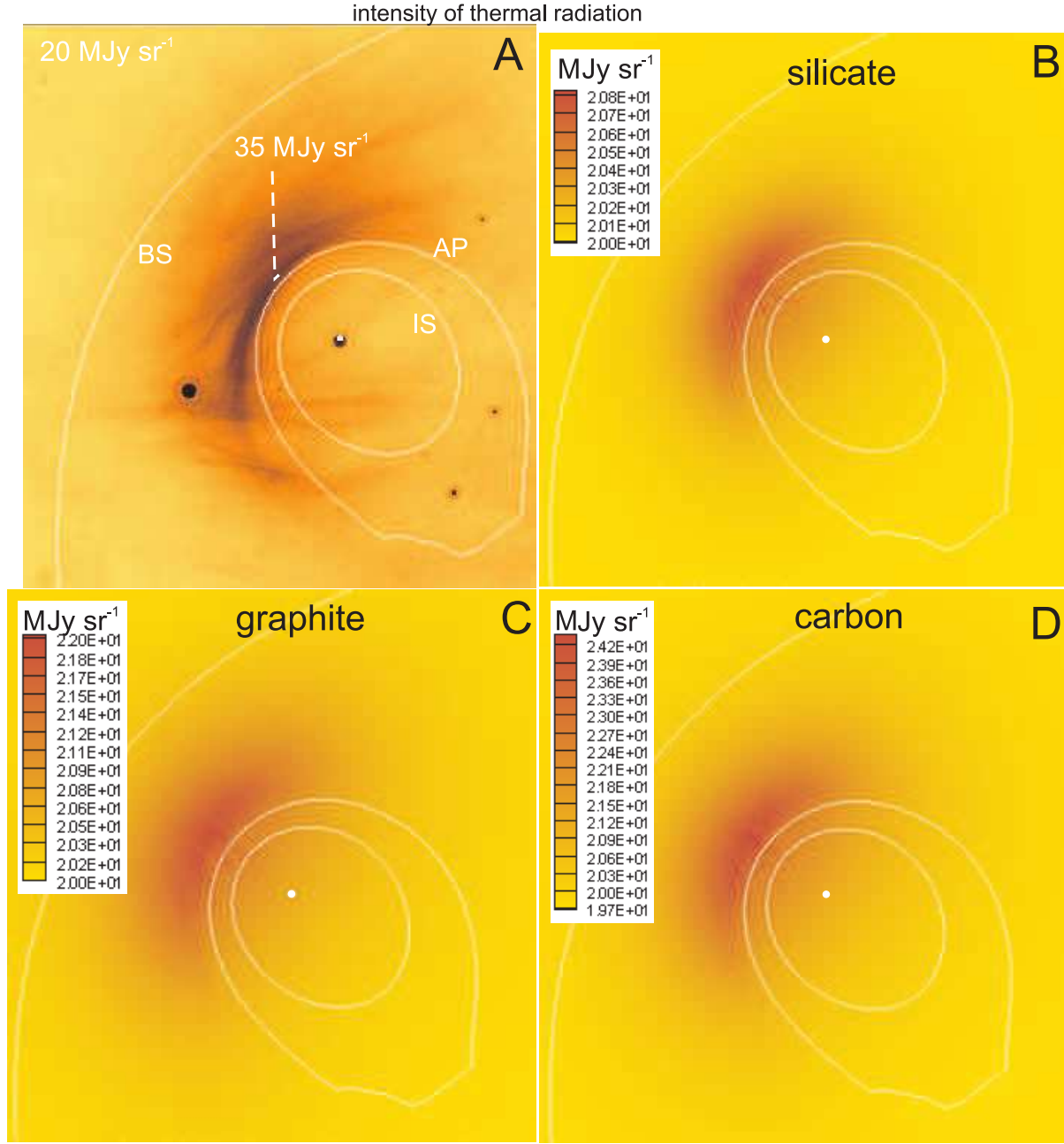}
\centering
\caption{A. MIPS 24~$\mu$m image of the astrosphere around $\kappa$\,Cas with white lines indication the
position of discontinuities (IS, AP and BS) in the model astrosphere.
B--D. Synthetic intensity maps of the thermal dust emission at $24\,\mu$m for silicates,
graphite and pure carbon. The model results are obtained for dust with the MRN size distribution with the
range of grain sizes of $0.2-3\, \mu$m and the dust temperature calculated from the local energy balance equation.
}
\label{fig:intens_tot}
\end{figure*}

Fig.~\ref{fig:intens_tot} shows the MIPS\,24~$\mu$m image of the astrosphere around $\kappa$\,Cas along with synthetic
maps of emission from silicate, graphite and pure carbon dust grains at the same wavelength. All intensities are given
in units of MJy sr$^{-1}$. Note that we added a constant intensity of 20~MJy sr$^{-1}$ to all synthetic maps to mimic
the background emission that is seen in the data. It is seen that for all types of dust grains there is a maximum
of intensity at the nose part of the astrosphere close to the astropause.
However, no separate filaments are visible. This is explained by two effects. The first one is the same as was discussed
above for the distribution of the dust number density -- no filaments can be distinguished for the mixture of dust grains
with the MNR size distribution. For the intensity maps this effect is even more pronounced than for the number density.
The reason is that small dust grains are more heated than the larger ones and therefore their contribution to the total
emission intensity is larger. In our model the filaments are clearly seen for grains with radii $r_{\rm d}=1-2\, \mu$m.
But these large grains are too cool to contribute to the total intensity maps. Note that the temperature of carbon grains
is much higher than the temperature of silicon and graphite ones (Fig.~\ref{fig:Td}), but this is still not enough
to make filaments visible because the efficiency absorption factor $Q_{\nu}$ at 24\,$\mu$m for carbon is much smaller than
that for graphite and silicon (see Fig.~\ref{fig:Qnu}).

The second effect is connected with the distribution of the dust temperature, which decreases with distance from the star
because the dust grains are heated mostly by the stellar radiation. Correspondingly, the intensity of the thermal dust
emission, which is proportional to the Planck function, is a strong function of temperature and hence of the distance
from the star. As a result, the thermal dust emission near the AP is much stronger than near the BS.

Fig.~\ref{fig:intens_tot} also shows that the emission intensity ratio of the brightest arc of the astrosphere
to the background is about an order of magnitude smaller compared with the observations.

\begin{figure*}
\includegraphics[scale=0.8]{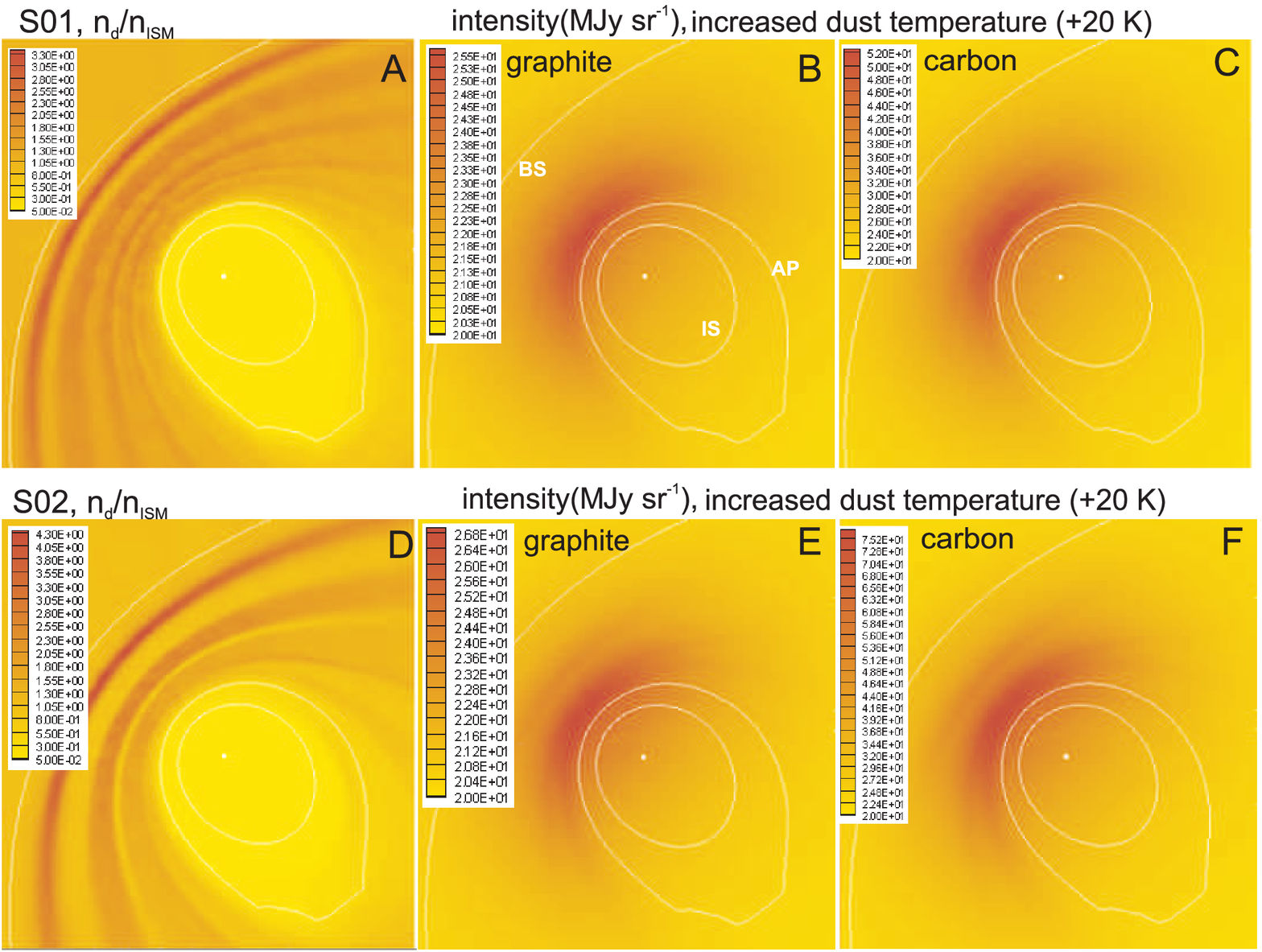}
\centering
\caption{Model results obtained for two narrow ranges of grain radii (SO1: $r_d\in[1.3,1.7]\, \mu$m and SO2: $r_d\in[1.8,2.2]\, \mu$m).
Plots A and D present dust number density in dimensionless units. Plots B, C, E and F present intensity maps for graphite and carbon.
The results are obtained with the dust temperature increased by 20~K everywhere. Discontinuities are shown by white lines.
}
\label{fig:intens_part}
\end{figure*}

Thus, we found that the observed filamentary structure cannot be explained in the framework of the model with
the classical MRN size distribution.
The actual dust size distribution however could differ from the MRN one. For example, Wang, Li \&
Jiang (2015) reported that the existence of
the very large ($0.5-6 \, \mu$m) dust grains in the ISM is confirmed by several independent observational evidences
(see also Lehtinen \& Mattila 1996; Pagani et al. 2010; Steinacker et al. 2015). Moreover, Wang et al. (2015) noted that ``if a
substantial fraction of interstellar dust is from supernova condensates, then $\mu$m-sized grains may
be prevalent in the ISM''.

Assuming that the large dust grains are indeed prevail in the local ISM, we examine the range of
dust parameters for which one can reproduce the observed filamentary structure of the astrosphere
around $\kappa$\,Cas. In particular, we assumed that the dust in the local ISM is composed
only of big grains (with the gas to dust mass ratio of 100) and performed calculations for two
narrow ranges of grain sizes with uniform distribution inside each range: $r_{\rm d}\in[1.3,1.7]\, \mu$m
and $r_{\rm d}\in[1.8,2.2]\, \mu$m; hereafter ranges SO1 and SO2, respectively.
Figs~\ref{fig:intens_part}~A and \ref{fig:intens_part}~D plot the distributions of the dust number
density in the observational plane for graphite and carbon grains with radii in the above two ranges.
One can discern five and three filaments, respectively, for SO1 and SO2. The filaments are wider than
in Fig.~\ref{fig:ndust} because now we consider a range of grain radii, while Fig.~\ref{fig:ndust}
was obtained for grains of a particular radius. We also calculated corresponding intensity maps, but
the filaments do not appear on them because of the low dust temperature, which rapidly decreases
with distance from the star (we do not show these maps since they are very similar to those
shown in Fig.~\ref{fig:intens_tot}). The temperature obtained as a solution of the local
energy balance (see Appendix~A) is high enough to produce filaments in the intensity maps at 24\,$\mu$m
only in a narrow region close to the astropause, while at larger distances the dust emission at this
wavelength is rapidly deceases.

We speculate that the dust grains might be hotter due to some additional heating processes.
To check how this will affect the intensity maps, we artificially increased the dust temperature
for both graphite and carbon grains by 20~K everywhere in the astrosphere. The resulting emission maps are presented
in Figs~\ref{fig:intens_part}~B--C and \ref{fig:intens_part}~E--F for both ranges of grain sizes
SO1 and SO2. One can see that with the
increase of the dust temperature the filaments become more pronounced. Qualitatively, these intensity
maps are quite similar to the MIPS image of the astrosphere around $\kappa$\,Cas. The absolute values
of the emission intensity are also similar to the observed ones (recall that the calculated
intensity is accurate within a factor of few due to the uncertainty in the ISM plasma density estimate).
We note also that the dust density maximum visible near the BS in the panels\,A and D of Fig.~\ref{fig:intens_part}
is absent in the intensity maps. This is again because of small dust temperature in this remote
part of the astrosphere.

Finally, we note that the smaller the dust grains the higher their temperature (see the panel\,A in
Fig.~\ref{fig:Td}), which implies that one can avoid the artificial increase of the dust temperature
if one adopts smaller dust grain radii. On the other hand, to make the filaments observable,
one needs to keep the same separation between them, which for the given dust grain radius is inversely
proportional to the magnetic field strength and the dust surface potential (see equation\,(\ref{eqn:Dgyr})).
From this it follows that the decrease of the grain size should be compensated by decrease of
$B$ or $U_{\rm d,ISM}$ (or both). A strong decrease of the magnetic field strength, however, is less appropriate
because, as discussed above, this would lead to the collapse of the outer shock region. The surface
potential of the dust grains is, in principle, a free parameter of the model and many different processes
can affect its value. If one adopts a factor of 10 smaller potential, then to  produce the same number
of filaments the radii of the dust grains could be a factor of $\approx3$ smaller than those adopted in
our modelling. Such grains are hot enough to produce observable filaments in the intensity maps.

\begin{figure*}
\includegraphics[scale=0.8]{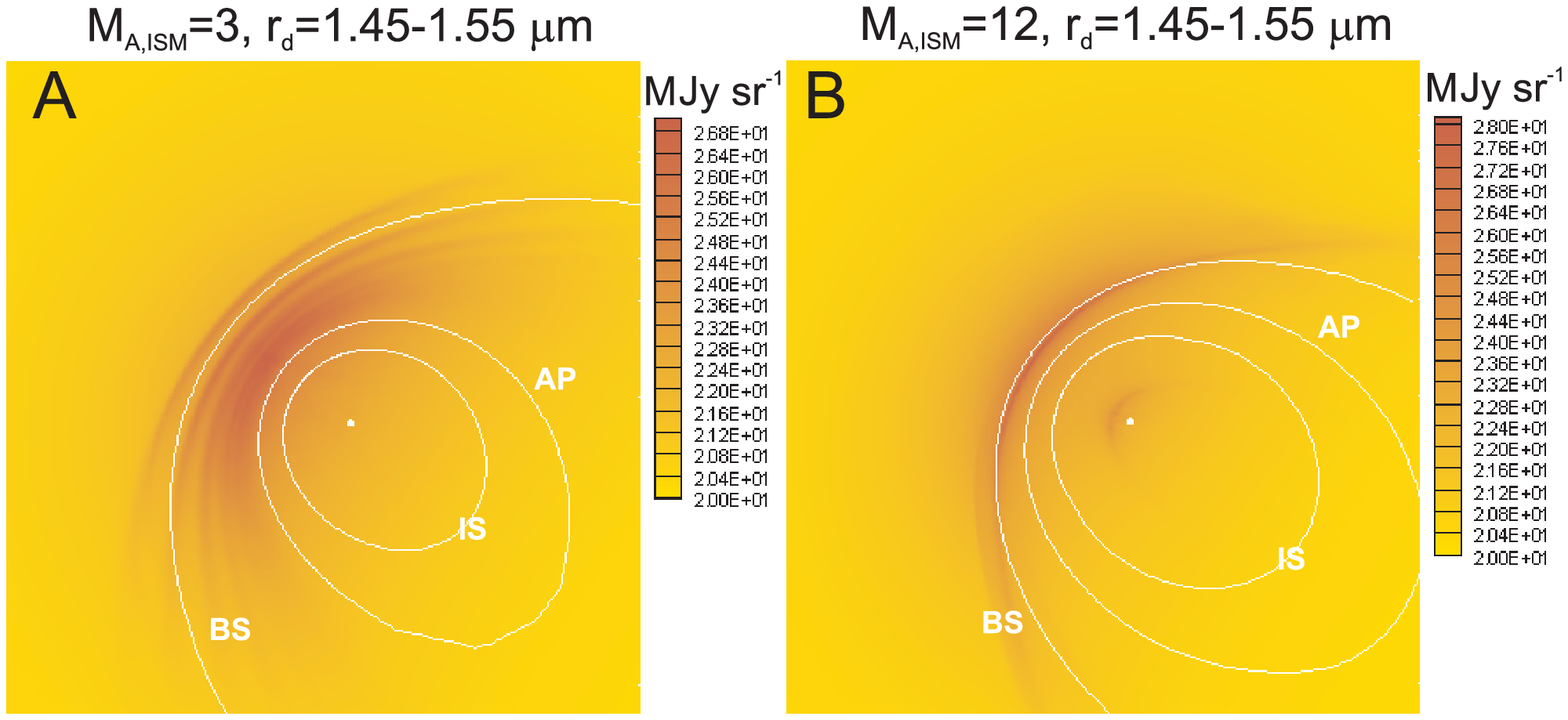}
\centering
\caption{Maps of the thermal dust emission calculated for graphite dust grains with radii
in the range $r_{\rm d}=1.45-1.55 \, \mu$m and temperatures artificially increased by 20\,K.
Plots in panels\,A and B correspond to models with $M_{\rm A,ISM}$=3 and 12, respectively.
$M_{\rm ISM}=3$ in both models.
}
\label{fig:dif_Ma}
\end{figure*}

\begin{figure*}
\includegraphics[scale=0.8]{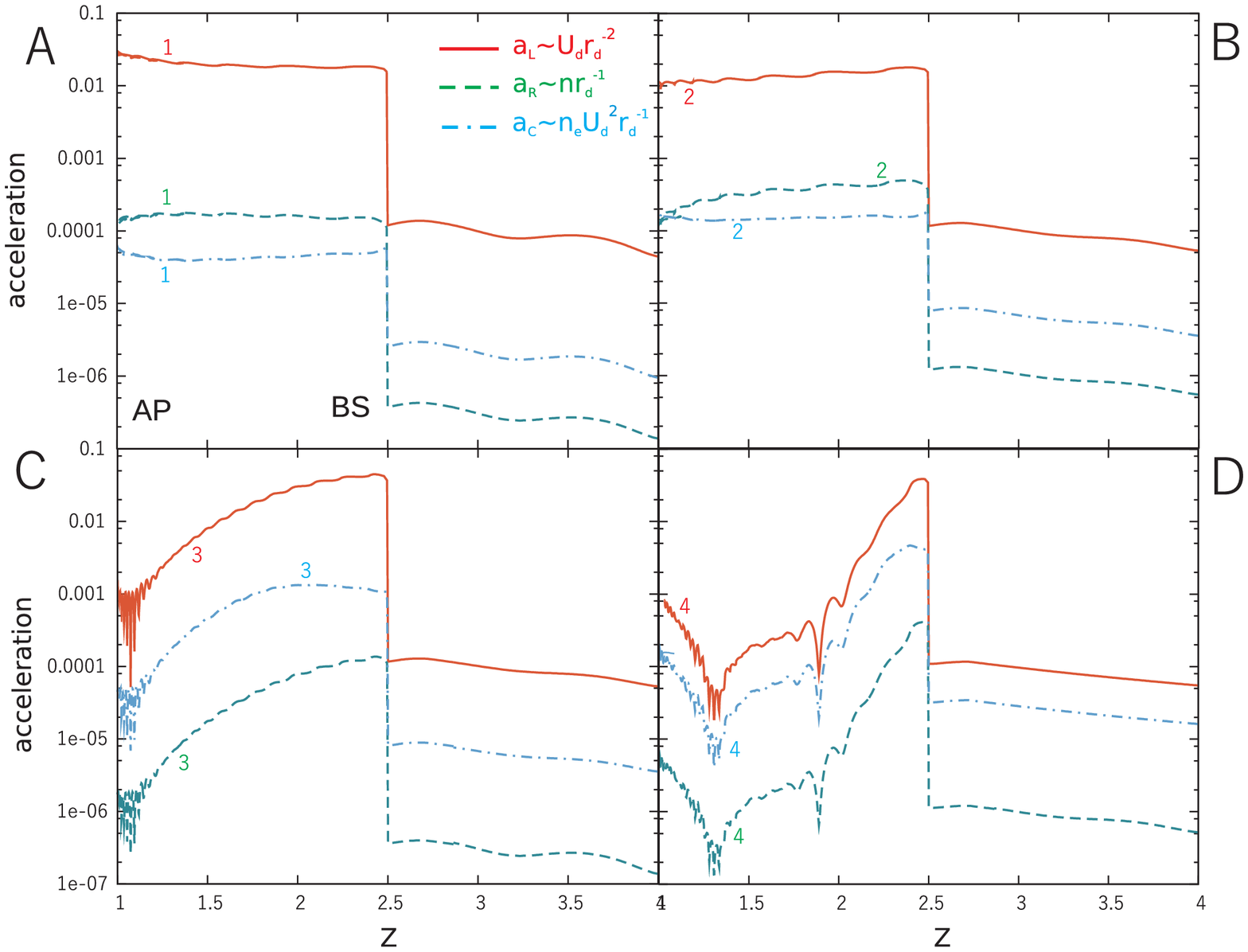}
\centering
\caption{Dimensionless acceleration ($F/m_{\rm d}$ normalized to $180 V_{\rm ISM}^2/R_{\rm obs}$) of a dust grain (with radius of 1\,$\mu$m) along its trajectory caused by the
Lorentz force (red curves), the direct drag force (green curves) and the Coulomb drag force (blue curves).
The results presented in panels A--D were obtained, respectively, for the following four pairs of values of the proton number
density and initial dust surface potential: 1) $n_{\rm p,ISM}=3 \, {\rm cm}^{-3}, U_{\rm d,ISM}$=0.75~V; 2) $n_{\rm p,ISM}=10
\, {\rm cm}^{-3}, U_{\rm d,ISM}$=0.75~V; 3) $n_{\rm p,ISM}=3 \, {\rm cm}^{-3}, U_{\rm d,ISM}=-1.3$~V, and
4) $n_{\rm p,ISM}=10 \, {\rm cm}^{-3}, U_{\rm d,ISM}=-$1.6~V.
}
\label{fig:acc}
\end{figure*}

\section{Summary and discussion}
\label{sec:dis}

In this paper, we performed 3D MHD numerical modelling of the astrosphere around $\kappa$\,Cas
in order to produce a synthetic map of its thermal dust emission at 24\,$\mu$m and to explain its
filamentary structure. We found that distinct filaments would appear in the emission map only if quite
large ($\mu$m-sized) dust grains are prevalent in the local ISM. The filamentary structure is not seen
for continuous power-law size distribution of dust because individual filaments merge with each other due
to the influence of small grains. Our model with large (1.3-2.2~$\mu$m) graphite and pure carbon dust grains
reproduces the observational data quite well, if the temperature of these grains in the region
where the filaments are formed is about 40 and 75~K, respectively. Comparison of the observed distance from
the star to the brightest arc in the astrosphere with the model results allows us to estimate the ISM number
density to be $3-11 \, {\rm cm}^{-3}$. We also constrain the local interstellar magnetic field strength to be
$18-35 \, \mu$G, which exceeds the typical field strength in the warm phase of the ISM (Troland \& Heiles 1986;
Harvey-Smith et al. 2011).

We performed test calculations with larger Alfvenic Mach numbers $M_{\rm A,ISM}$=3 and 12 that corresponds to weaker interstellar magnetic field
 (10 and 2.7~$\mu$G, respectively, for $n_{\rm p,ISM}$=3~cm$^{-3}$). The synthetic maps of thermal dust emission for graphite dust
grains with radii of $1.45-1.55 \,\mu$m are presented in Fig.~\ref{fig:dif_Ma}. In this figure the filaments are
more distinct compared to, e.g., Fig.~\ref{fig:intens_part} because the more narrow range of grain radii
is considered. For smaller $B_{\rm ISM}$ the outer shock layer (confined between the BS and the AP) becomes
thinner. The model with $M_{\rm A,ISM}$=3 is still appropriate: the filaments are seen and the emission
intensity is close to the observed one. Note that the outermost filament in the panel\,A
of Fig.~\ref{fig:dif_Ma} is in fact located between the AP and the BS, and appears beyond the BS because of
the projection effect. The model with $M_{\rm A,ISM}$=12 provides a too small separation between the AP and
the BS, so that the filaments merge with each other and the model cannot reproduce the observations for any
grain size distribution. It is also interesting to note that there is a small arc close to the star in the
case of weak magnetic field ($M_{\rm A,ISM}$=12). This arc is formed by large dust grains penetrating inside
the IS. Near the star these grains are swept out by the stellar radiation and appear as an arc.

Our numerical calculations are performed under assumption of constant dust charge.
In general, the grain charge is determined by the balance between three main processes: impinging
of protons and electrons, secondary electron emission due to electron impacts
(this is especially important for hot plasma with $T\ga10^5$\,K) and photoelectron emission caused by
the external interstellar and stellar radiation. We performed estimations of the changes of the dust charge
and found that they are not more than 30 per cent in the considered region between the BS and the AP.
Therefore we can neglect them and assume that the dust grain charge does not vary along the trajectory.

In our calculations, we neglected the drag forces caused by direct collisions of the dust
grains with ions and electrons (direct drag force), and by the electromagnetic Coulomb interaction
(Coulomb drag force), although our numerical
model allows us to take them into account. Ochsendorf et al. (2014) found that the Coulomb drag force
can affect the formation of so-called ``dust waves'' -- arclike enhancements of dust density
around weak-wind stars -- created because of decoupling of the dust grains from the gas by stellar radiation
force. Namely, they showed that inclusion of the Coulomb drag in the model leads to a strong
dust-gas coupling, which prevents the formation of the dust waves. To clarify whether or not the drag
forces could be important in our calculations, below we discuss the model parameters
which determine their relative contributions to the dust motion in the astrosphere.
Fig.~\ref{fig:acc} shows accelerations ($a=F/m_{\rm d}$) of a dust grain (with radius $r_{\rm d}=1 \, \mu$m)
along its trajectory due to three forces: the Lorentz force ($a_{\rm L}$), the direct drag force
($a_{\rm R}$) and the Coulomb drag force ($a_{\rm C}$). Note that $a_{\rm L}\sim q/m_{\rm d} \sim
U_{\rm d}/r_{\rm d}^2$, $a_{\rm R}\sim n_{\rm p}/r_{\rm d}$, and $a_{\rm C}\sim n_{\rm p} U_{\rm d}^2/r_{\rm d}$.
These relations determine the balance between different forces for the chosen model parameters. Accelerations are
calculated for the following four pairs of values of the proton number density and the initial dust surface potential:
$n_{\rm p,ISM}=3 \, {\rm cm}^{-3}, U_{\rm d,ISM}=0.75$~V (hereafter, case\,1; see panel A in Fig.~\ref{fig:acc}),
$n_{\rm p,ISM}=10 \, {\rm cm}^{-3}, U_{\rm d,ISM}$=0.75~V (case 2; panel B), $n_{\rm p,ISM}=3 \, {\rm cm}^{-3},
U_{\rm d,ISM}=-1.3$~V (case 3; panel C), and $n_{\rm p,ISM}=10 \, {\rm cm}^{-3}, U_{\rm d,ISM}=-1.6$~V (case 4; panel D).
It is seen from Fig.~\ref{fig:acc} that in cases\,3 and 4 the Coulomb drag force is larger than the direct drag force
because of the large potential of the dust grains. One can see also that the Lorentz force is by one-two orders of magnitude
larger than the drag forces in cases\,1--3 and becomes comparable to the Coulomb drag force in case\,4.
Thus, it is seen that for the parameters considered in our paper ($n_{\rm p,ISM}=3-11 \, {\rm cm}^{-3}$ and $U_{\rm d,ISM}$=0.75~V)
the influence of the drag forces can be safely neglected. However, for higher gas densities and/or dust potentials
their effect could be significant.

Our explanation of the cirrus-like structure of the astrosphere around $\kappa$\,Cas implies that
the thermal mid-IR emission of the dust originates in regions spatially separated from the region of
the bulk optical line emission. Also, we expect that in the optical wavelengths the astrosphere
should have a smooth appearance, unless it is deformed by (magneto)hydrodynamic instabilities.
But even in this case, the optical filaments should not correlate with the mid-IR cirrus-like ones.
In principle, the difference between the mid-IR and optical appearances of an astrosphere could be detected,
provided that it is nearby enough to allow us to resolve the layer between the astropause and
the bow shock. Unfortunately, with the existing optical surveys we were not able to detect the optical
counterpart of the astrosphere of $\kappa$\,Cas, particularly because this bright ($V\approx4$ mag) star
outshines all around it. Optical imaging with narrow-band filters could potentially be of value in detection
of the astrosphere and in verifying our model.

We realize that the mechanism for origin of the filamentary structure of astrospheres is not unique
and that the observed filaments might be produced by various different processes.
For example, they could arise because of the rippling effect in radiative shocks, i.e. due
to variations in the projection of the shock velocity along the line of sight (Hester 1987).
Also, the filaments could originate because of time-dependent variations of the wind velocity
(Decin et al. 2006) or might be caused by instabilities in the bow shock and contact discontinuity
(Dgani, Van Buren \& Noriega-Crespo 1996). Numerical simulations by many authors (e.g.,
van Marle et al. 2011, 2014; Mackey et al. 2012; Meyer et al. 2014a,b; Acreman et al. 2016) indeed
show that astrospheres may
be subject to various types of instabilities. However, it is a challenge to separate the real instabilities
from the numerical ones, and we refrain in this paper from discussing this problem in depth.

\begin{figure}
\includegraphics[scale=0.45]{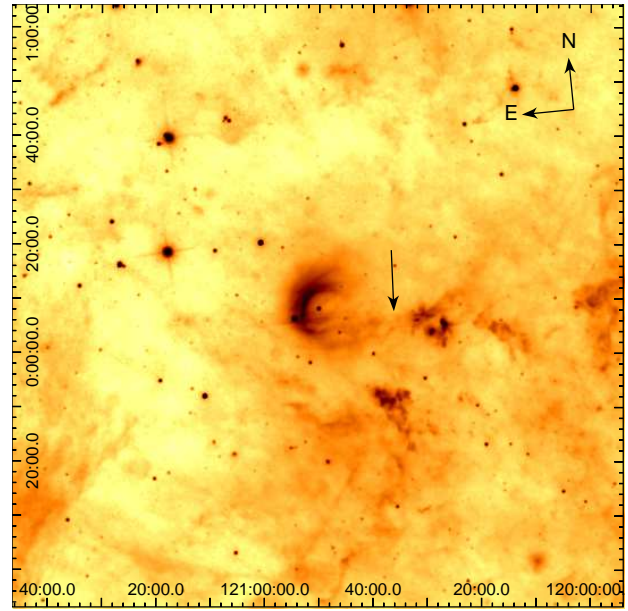}
\centering
\caption{$1\degr\times1\degr$ {\it WISE} 22~$\mu$m image of the field containing $\kappa$\,Cas
and its astrosphere. The pillar to the west of $\kappa$\,Cas is indicated by an arrow (see text for details).
The coordinates are the Galactic longitude and latitude on the horizontal and vertical scales, respectively.
At a distance of 1 kpc, $1\degr$ corresponds to $\approx17$ pc.}
\label{fig:WISE}
\end{figure}

To conclude, we note that our model does not explain the origin of the almost straight filaments in the south part of the
astrosphere, which intersect the brightest arc at almost right angle (see Fig.\,\ref{fig:kapCas}). These filaments might be due
to much more complex structure of the interstellar magnetic field than assumed in our modelling (cf. Gvaramadze et al. 2011b).
Also, inspection of the {\it WISE} 22\,$\mu$m image of the region around $\kappa$\,Cas shows that these filaments have the same
orientation as an elongated pillar to the west of the star (see Fig.\,\ref{fig:WISE}), which points to the possibility that
their origin might be due to interaction of the astrosphere with the inhomogeneous local ISM. Modelling of this interaction
is, however, beyond the scope of the present paper.

\section*{Acknowledgments}
This work is supported by the Russian Science Foundation grant No.~14-12-01096
and is based in part on archival data obtained with the {\it Spitzer Space Telescope},
which is operated by the Jet Propulsion Laboratory, California Institute of Technology
under a contract with NASA, and has made use of the NASA/IPAC Infrared Science Archive,
which is operated by the Jet Propulsion Laboratory, California Institute of Technology,
under contract with the National Aeronautics and Space Administration, the SIMBAD
data base and the VizieR catalogue access tool, both operated at CDS, Strasbourg, France.

\appendix

\section{Calculation of dust temperature}

The dust temperature in the astrosphere can be found by solving the dust thermal balance for equilibrium.
This approach is justified because the heating and cooling time-scales are shorter than the
characteristic time-scale of dust grain motion in the astrosphere. We do not consider the radiative transfer
and multiple scattering of photons because of small dust number density that allows us to use an optically
thin approximation. The dust grains are heated due to absorption of stellar and interstellar radiation, while
their cooling is caused mostly by thermal black body emission. The energy balance for a single dust grain can
be stated as follows (see, e.g., Hocuk et al. 2017):
\begin{eqnarray}
4\pi\,r_{\rm d}^2\,\int_0^{\infty}Q_{\nu}B_{\nu}(T_{\rm d})\,d\nu=r_{\rm d}^2 \, \int_0^{\infty} Q_{\nu}F^*_{\nu}(r,T_*)\,d\nu+ \\
 +2\pi\,r_{\rm d}^2\,\int_0^{\infty} Q_{\nu}J_{\nu}\,d\nu \nonumber,
\label{eq:Td}
\end{eqnarray}
where $F^*_{\nu}=\pi\, (R_*/r)^2 B_{\nu}(T_*)$ is the stellar radiation at point $r$, $J_{\nu}$ is the intensity of the isotropic
stellar radiation field (see fig.~2 in Hocuk et al. 2017). $R_*$ and $T_*$ for $\kappa$\,Cas are given in Table~\ref{tab:sum}.
Equation~(\ref{eq:Td}) is solved numerically and $T_{\rm d}$ is found for any grain radii $r_{\rm d}$ and distances $r$.

\section{Dust number density in the ISM}

In the case of power-law size distribution, the ISM number density of dust grains with
radius of $[r_d-dr_d/2; r_d+dr_d/2]$ is $dn_{\rm ISM}(r_{\rm d})=N_{\rm ISM}r_{\rm d}^{-3.5} d r_{\rm d}$
and their mass density is $d \rho_{\rm d,ISM}=m_{\rm d} \, dn_{\rm d,ISM}(r_{\rm d})$.
$m_{\rm d}=4/3\pi \, r_{\rm d}^3 \, \rho_{\rm d}$ is the mass of a dust grain,
$\rho_{\rm d}=2.5 \, {\rm g} \, {\rm cm}^{-3}$ is a typical mass density for the
interstellar grain material. We assume that the gas to dust mass ratio in the ISM is about
100. The mass density of the dust in the ISM is:
\[
 \rho_{\rm d,ISM}=\frac{4}{3}\pi \rho_{\rm d} N_{\rm ISM} \int_{r_{\rm d,min}}^{r_{\rm d,max}}
 r_{\rm d}^{-0.5} dr_{\rm d}.
 \]
For the gas to dust mass ratio of $\rho_{\rm p,ISM}/\rho_{\rm d,ISM}=100$, one has
\[
   N_{\rm ISM}=\frac{3}{800 \pi}  \frac{\rho_{\rm p,ISM}}{\rho_{\rm d}} \frac{1}{r_{\rm d,max}^{0.5}-r_{\rm d,min}^{0.5}}.
\]
Note that $[N_{\rm ISM}]={\rm cm}^{-0.5}$.

\end{document}